\documentclass{article}
\usepackage{spconf,amsmath,graphicx,hyperref,xcolor}
\usepackage{cite}
\usepackage{booktabs} 
\usepackage{colortbl}
\usepackage[dvipsnames]{xcolor}
\usepackage{enumitem}
 \title{Target speaker anonymization in multi-speaker recordings}
\name{Natalia Tomashenko$^1$, Junichi Yamagishi$^2$, Xin Wang$^2$, Yun Liu$^2$, Emmanuel Vincent$^1$
\thanks{This work was done in the context of the 
Inria-NII TrustedSpeech Associate Team and was  
supported by the French National Research Agency under
project Speech Privacy and project IPoP of the Cybersecurity PEPR. Experiments were carried out using the Grid’5000 testbed.}
\address{$^1$\textit{Universit\'e de Lorraine, CNRS, Inria, Loria, F-54000}, Nancy, France  \\
$^2$\textit{National Institute of Informatics, 2-1-2 Hitotsubashi, 101-8430}, Tokyo, Japan}}
\begin{document}
\ninept
\maketitle
%
% \vspace{-3pt}

\begin{abstract}
%   The abstract should contain about 100 to 150
% words, 

Most of the existing speaker anonymization research has focused on single-speaker audio, leading to the development of techniques and evaluation metrics optimized for such condition. This study addresses the significant challenge of speaker anonymization within multi-speaker conversational audio, specifically when only a single target speaker needs to be anonymized. This scenario is highly relevant in contexts like call centers, where customer privacy necessitates anonymizing only the customer's voice in interactions with operators.
Conventional anonymization 
methods are often not suitable for this task. 
Moreover, current evaluation 
methodology
does not allow us to accurately assess  privacy protection and  utility in this complex multi-speaker scenario.
This work aims to bridge these gaps by exploring effective strategies for targeted speaker anonymization in conversational audio, highlighting potential problems in their development and proposing corresponding improved evaluation methodologies\footnote{Audio samples are available at \url{https://sites.google.com/view/target-speaker-anonymization}}.

\end{abstract}
\vspace{-1pt}

\section{Introduction}
\label{sec:intro}
\vspace{-3pt}

The extensive deployment of spoken language technologies raises substantial privacy concerns and development of regulatory frameworks such as the European General Data Protection Regulation (GDPR) \cite{nautsch2019gdpr}. Voice recordings is a rich source of personal information; beyond merely identifying an individual, speech data contains sensitive attributes like age, gender, health status, emotional state, personality traits, ethnic origin, and socioeconomic standing. 
To address the privacy challenges, voice anonymization has emerged as a critical and prevalent methodology for protecting the privacy of speech data. Anonymization aims at suppressing the personally identifiable traits of a speaker,  while preserving the linguistic and some  paralinguistic content \cite{tomashenko2020introducing}.
Voice anonymization techniques 
include
two broad categories of methods:
  signal processing-based methods \cite{patino2020speaker,mawalim22_spsc,gupta2020designn,tavi2022improving}, that employ simple signal transformations to alter voice characteristics; and neural voice conversion methods \cite{fang2019speaker,miao2023speaker,srivastava2021,champion2023,yao2024musa,miao2022language,yao2024npu,saini2023speaker,webber2024voice}, that operate by disentangling various speech attributes --- such as content, speaker characteristics, pitch, and emotion --- before selectively anonymizing specific attributes and reconstructing the speech signal using speech synthesis models.

 Previous research on speaker anonymization has focused primarily on single-speaker audio scenarios, which has shaped the development of anonymization techniques and evaluation metrics specifically tailored to these contexts.
 In contrast, this study addresses the challenge of speaker anonymization in conversational audio, specifically focusing on scenarios where only one particular speaker within a multi-speaker dialogue needs to be anonymized.  This scenario is particularly common in call centers, where conversations between customers and operators are recorded, but only the customer's voice needs to be anonymized to protect their privacy.
Conventional anonymization methods fall short in this scenario as they cannot selectively target individual speakers within conversations. 
 Additionally, current evaluation metrics are insufficient for accurately measuring both privacy protection and audio utility in these specific contexts.
The only existing work related to
multi-speaker anonymization \cite{miao2025benchmark} proposes an initial attempt to provide a multi-speaker anonymization benchmark,
however, while  the potential privacy leakage caused by overlapping segments is acknowledged  by the authors, the proposed practical  solution  
is limited to  non-overlapping conversations. 
More realistic and challenging scenarios include conversations with 
interfering speech of multiple speakers \cite{cosentino2020librimix} that is the main focus of the current research.

The key contributions of this work 
are the following: (i) introduction of a novel target speaker anonymization framework, specifically designed to address privacy preservation challenges within multi-speaker conversations;
(ii) development of an  evaluation methodology for assessing both privacy protection and utility preservation in anonymized speech;
(iii) experimental investigations leveraging two state-of-the-art target speaker extraction methods and an  anonymization technique, analyzing their performance across diverse overlapped speech conditions and the inherent limitations and challenges of the proposed models.

       \begin{figure*}
        \centering
        \includegraphics[width=0.92\linewidth]{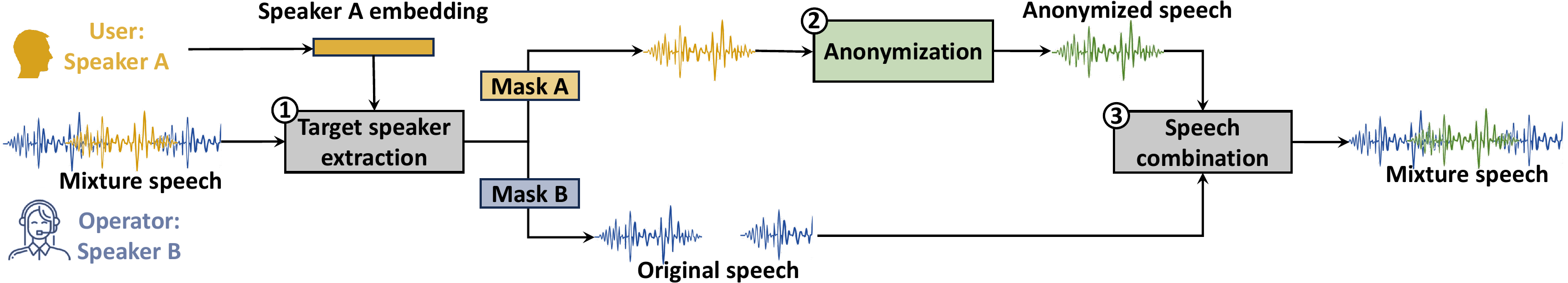}
        \caption{Target speaker anonymization (TSA)}
        \label{fig:tsa}
    \end{figure*}

\vspace{-4pt}
\section{Target speaker anonymization}
\label{sec:tsa}
\vspace{-4pt}

 To address the challenges of privacy protection in multi-speaker conversations, we propose a novel framework called \textit{target speaker anonymization (TSA)} for selectively anonymizing a specific speaker in multi-speaker recordings. 
To achieve speaker anonymization for only the target speaker, we propose a pipeline approach that combines target speaker extraction (TSE) \cite{wesep_2024} with speaker voice anonymization techniques of only the designated target speaker. 

TSE focuses on isolating the speech of the designated target speaker from overlapped multi-talker speech --- a common challenge often referred to as the ``\textit{cocktail party problem}". By accurately extracting the target speaker's voice, subsequent anonymization steps can be applied effectively without affecting other speakers in the recording.
The overview of the TSA is illustrated 
in Fig.~\ref{fig:tsa} for a common scenario for a two-speaker recording that features a user (designated as \textit{Speaker A}) and a call center operator (designated as \textit{Speaker B}).
In this illustrative example, the user (\textit{Speaker A}) is identified as the target speaker, meaning their voice will undergo anonymization to protect their personal identity. The operator (\textit{Speaker B}), conversely, is a non-target speaker whose speech will remain in its original, non-anonymized form to maintain operational clarity and context.

TSA includes three main steps: 
(1)~Target speaker extraction (TSE).  TSE typically involves estimating a complex-valued soft mask (denoted as \textit{Mask A} in Fig.~\ref{fig:tsa}) for the time-frequency spectral bins containing the target speaker, using a pre-provided neural speaker embedding vector. This technique 
    isolates the target speaker from the mixture, allowing us to perform speaker anonymization exclusively on the extracted speech segments of \textit{Speaker~A}.
   % \item
(2)~Anonymization of the voice of the target \textit{Speaker~A}.
   % \item
(3)~Combination of the anonymized speech waveform of the target speaker and the (residual) speech waveform of \textit{Speaker~B} which remains non-anonymized.

By effectively isolating the target speaker from the mixed audio at \textit{Step 1}, we ensure that their speech is cleanly extracted. This allows us to then perform speaker anonymization at \textit{Step 2} exclusively on these extracted segments, maintaining privacy for the target individual without affecting other voices or sounds in the conversation.

\vspace{-3pt}

\section{Evaluation methodology}
\label{sec:eval_met}

\vspace{-3pt}

  This section introduces enhanced evaluation metrics designed to specifically measure both privacy protection and utility preservation in target speaker-aware anonymization scenarios.

\vspace{-3pt}

\subsection{Privacy assessment}

    Beyond integrating target speaker extraction and anonymization techniques, we must improve evaluation metrics for privacy protection. Traditionally, privacy has been assessed by confirming that anonymized speech cannot be linked to the original speaker using equal error rate (EER) with automatic speaker verification (ASV) models \cite{tomashenko2020introducing} that are considered as attack systems. This evaluation remains crucial when the extracted speech contains only the target speaker.
    However, imperfections in target speaker extraction may leave residual speaker information in time-frequency bins that aren't properly masked. Attackers could potentially exploit these residuals to re-identify the original speaker.

\subsection{Utility assessment}

   Evaluating the utility of anonymized speech is crucial alongside privacy protection. Regarding utility, it is essential that anonymized speech remains functional for downstream tasks.  A primary concern is ensuring that anonymized speech remains intelligible when mixed with other speakers, allowing accurate transcription by automatic speech recognition (ASR) systems. 
     Additionally, we must ensure that anonymized speech within multi-speaker recordings can still support effective speaker diarization. This capability is essential in practical applications like call center analytics, where distinguishing between different speakers  remains important. Our framework must therefore be evaluated both on ASR performance and compatibility with speaker diarization systems. These utility metrics collectively ensure that anonymized speech maintains its practical value for real-world applications while protecting the privacy of the target speaker.
\vspace{-3pt}
         
\subsubsection{Primary utility metric}

\label{sssec:primary_utility}

\vspace{-1pt}

Traditionally, in a single-speaker audio voice anonymization scenario, objective utility evaluation relies on the word error rate (WER) metric, which is obtained by computing decoding errors of the ASR system.
However, in multi-speaker recordings, the task becomes more intricate, as anonymization must not only preserve the content spoken by the target speaker but also maintain the intelligibility of other speakers and the overall conversational flow. Therefore, TSA requires a more comprehensive  evaluation metric.
As a primary utility metric to evaluate verbal content preservation  for all speakers in the mixture obtained by TSA, we consider a minimum-permutation word error rate (cpWER) \cite{watanabe20b_chime} obtained by speaker-attributed ASR \cite{polok2025dicow}.
The tcpWER is computed using the meeting transcription evaluation toolkit \textit{MeetEval}\cite{MeetEval23}\footnote{\url{https://github.com/fgnt/meeteval}} as follows:
(i) All utterances for each speaker are concatenated in both the reference and hypothesis files. (ii) The WER is then computed by comparing the reference to every possible speaker permutation of the hypothesis. (iii) The lowest WER among these permutations is selected, representing the optimal speaker alignment.
More specifically, we consider
time-constrained minimum-permutation word error rate (tcpWER), that is a variation of cpWER with incorporated temporal constraints into the  Levenshtein distance used for WER computation \cite{MeetEval23}.
By utilizing temporal information, the tcpWER prevents the matching of words that are temporally distant, thus offering a more accurate assessment of speech transcription quality.
Consequently, the tcpWER is directly and significantly affected by the accuracy of speaker diarization results.

\vspace{-3pt}

\subsubsection{Auxiliary utility metrics}
\label{sssec:auxil_utility}
\vspace{-1pt}

Beyond tcpWER, our
primary
utility metric, to evaluate the intermediate steps of the pipeline and gain deeper insights into overall performance and the influence of different system components, we analyze additional metrics: 
(i)
Diarization error rate (DER). Effective speaker diarization, which involves identifying ``\textit{who spoke when}", is a crucial utility metric. The anonymized audio must still allow for accurate segmentation of speech by speaker and correct assignment of speaker labels.
(ii) WER of the ASR system for anonymized speech signal of the target speaker (after \textit{Step 2} in Fig.~\ref{fig:tsa} and before combining speech signals).

\subsection{TSE quality assessment}
\label{ssec:tse-quality}
\vspace{-1pt}

In the considered privacy preservation scenario, both the user and  attacker should efficiently utilize TSE models to perform their respective tasks.
The quality of TSE impacts both the privacy and  utility of the anonymized speech, which is evaluated using privacy and utility metrics. As an additional intermediate metric for estimating TSE quality, we employ the 
scale-invariant signal-to-distortion ratio
SI-SDR  \cite{le2019sdr}. 
SI-SDR quantifies how accurately the separated speech signal matches the original reference signal, while inherently accounting for any scaling differences between them. A higher positive SI-SDR value signifies superior separation quality, indicating that the TSE model is more effective at isolating the target speaker's voice.

\vspace{-9pt}

\section{Experimental setup}
\label{sec:experiments}

\subsection{Evaluation data}
\label{ssec:data}

Experiments were conducted on the \textit{SparseLibri2Mix} subset of the open-source dataset
\textit{SparseLibriMix}\footnote{\label{libr-note}\url{https://github.com/popcornell/SparseLibriMix}} \cite{cosentino2020librimix}. 
This dataset was designed by combining recordings of two speakers from the \textit{LibriSpeech test-clean} subset to create more realistic, conversation-like scenarios, featuring 5 different overlapping versions (conditions): 
$20\%$, $40\%$, $60\%$, $80\%$, $100\%$  overlap.
This dataset includes 500 mixture files per condition 
($500\cdot 5=2500$ 
mixture files corresponding to about 5 hours of speech in total) from 40 speakers.
For our analysis, the first speaker in a mixture is designated as the target speaker, whose voice is to be anonymized and the second speaker is considered the non-target speaker, whose voice remains unchanged\footnote{First and second speakers in the mixtures are denoted as  $s1$  and $s2$ respectively in 
the \textit{SparseLibriMix} corpus release$^{\ref{libr-note}}$}.

For TSE experiments, an utterance (not present in the mixture files) from the \textit{LibriSpeech test-clean} dataset was randomly selected for a reference speaker in each mixture file to compute target speaker embeddings.
For ASV experiments, 15 utterances from the \textit{LibriSpeech test-clean} dataset were used for each speaker to compute enrollment speaker embeddings. The set of enrollment utterances was disjoint from the utterances used in the mixtures. Thus, the total number of trials per condition is $20000$: 
$500$ same-speaker trials and $19500$  different-speaker trials.

\subsection{Anonymization system}
\vspace{-1pt}

For experiments, as an example anonymization system, we use one of the strongest in terms of privacy  anonymization baseline systems from the VoicePrivacy 2024 Challenge \cite{tomashenko2024voiceprivacy}, that is based on
 using acoustic vector quantized bottleneck (VQ-BN) features extracted from an ASR acoustic model. 
 This system, denoted in \cite{tomashenko2024voiceprivacy} as B5, provides an enhanced disentanglement of speaker identity from linguistic content through vector quantization.
The pipeline extracts pitch and acoustic VQ-BN features. These features, combined with a designated speaker identity (represented as a one-hot vector corresponding to a speaker from the training dataset (\textit{train-clean-100 LibriTTS}), are then directly used to synthesize an anonymized speech waveform through a HiFi-GAN network. 
Data anonymization in all experiments was performed on the \textit{utterance level}.

\subsection{Target speaker extraction models}
\vspace{-1pt}

For our experiments, we utilized 
two 
distinct TSE models to investigate how the quality of separated speech impacts TSA performance. The two models were chosen in the preliminary experiments from multiple TSE model types \cite{liu2024libri2vox, wesep_2024} according to their performance in terms of 
the SI-SDR metric on the \textit{SparseLibriMix}~\cite{cosentino2020librimix} data.

\begin{table}[tb]
    \centering
    \resizebox{0.25\textwidth}{!}{
    \begin{tabular}{lcc}
        \toprule
         & \textbf{Original}  & \textbf{Anonymized}  \\
        \midrule
        EER,\% & 3.0 & 32.4 \\
        WER, \% & 2.7 & 6.0 \\
        \bottomrule
    \end{tabular}}
    \caption{Privacy and utility evaluation for a single-speaker scenario.}
    \label{tab:single-chan}
\end{table}

% \begin{table}[tb]
%     \centering
%     \resizebox{0.44\textwidth}{!}{
%     \begin{tabular}{lcccccc}
%         \toprule
%         \textbf{Overlap, $\%$} & \textbf{0} & \textbf{20} & \textbf{40} & \textbf{60} & \textbf{80} & \textbf{100} \\
%         \midrule
%         Conformer TSE & 24.1 & 17.9 & 15.8 & 14.6 & 14.0 & 14.0 \\
%         WeSep BSRNN TSE & 22.1 & 18.6 & 17.5 & 17.2 & 16.7 & 16.2 \\
%          \bottomrule
%     \end{tabular}}
%     \caption{Comparison of TSE models  at different overlap percentages in terms of SI-SDR.}
%     \label{tab:si-sdr}
% \end{table}

\begin{table}[tb]
    \centering
    \resizebox{0.375\textwidth}{!}{
    \begin{tabular}{lccccc}
        \toprule
        \textbf{Overlap, $\%$}  & \textbf{20} & \textbf{40} & \textbf{60} & \textbf{80} & \textbf{100} \\
        \midrule
        Conformer TSE  & 17.9 & 15.8 & 14.6 & 14.0 & 14.0 \\
        WeSep BSRNN TSE  & 18.6 & 17.5 & 17.2 & 16.7 & 16.2 \\
         \bottomrule
    \end{tabular}}
    \caption{Comparison of TSE models  at different overlap percentages in terms of SI-SDR.}
    \label{tab:si-sdr}
\end{table}

% 5.0
% 12.9

% 5.7
% 31.4
% 17.8
% 34.0
% 17.8
% 21.2
% 36.4

\begin{table*}[thb]
    \centering
    % \begin{tabular}{ccccccccc}
    \resizebox{0.89\textwidth}{!}{
    \begin{tabular}{lll|llllll|llllll}
        \toprule
        \textbf{Step} & \textbf{Metric,\%} & \textbf{Data}  & \textbf{20\%} & \textbf{40\%} & \textbf{60\%} & \textbf{80\%} & \textbf{100\%} & \textbf{Aver} &  \textbf{20\%} & \textbf{40\%} & \textbf{60\%} & \textbf{80\%} & \textbf{100\%} & \textbf{Aver}\\
\midrule
        Original  & tcpWER & orig & 4.3 & 4.3 & 4.5 & 4.6 & 7.2 & 5.0\\
         mixture & DER & orig  & 27.5 & 17.6 & 9.5 & 4.8 & 5.4 & 12.9 \\
\midrule
%  
% 4.6
% 31.6
% 14.4
% 23.9
% 14.6
% 16.9
% 36.9
& & & \multicolumn{6}{c|}{\textbf{Conformer TSE}} & \multicolumn{6}{c}{\textbf{WeSep BSRNN TSE}} \\
\midrule
         1. TSE$_{\text{user}}$ & EER & orig  & 4.8 & 5.2 & 5.6 & 6.4 & 6.2 & 5.7		& 4.4 &	4.2	& 4.8	& 4.8 &	4.8	& 4.6 \\
          & WER & orig  & 20.7 & 18.6 & 17.6 & 16.4 & 15.6 & 17.8	 &	21.3 & 	14.6	& 12.3	& 12.3	& 11.3 &	14.4 \\ 
        2. TSA& EER & anon  & 31.8 & 31.4 & 31.6 & 30.6 & 31.6 & 31.4	 &	33.0	& 31.2 & 	31.4 &	31.4 &	31.2 &	31.6\\ 
           & WER & anon  & 41.7 & 35.8 & 33.0 & 29.9 & 29.6 & 34.0 & 	31.4 & 	25.7 & 	21.6 & 	21.6 & 	19.0 & 	23.9 \\
        % 2. TSA & WER & orig  & 20.7 & 18.6 & 17.6 & 16.4 & 15.6 & 17.8	 &	21.3 & 	14.6	& 12.3	& 12.3	& 11.3 &	14.4 \\ 
        %  & WER & anon  & 41.7 & 35.8 & 33.0 & 29.9 & 29.6 & 34.0 & 	31.4 & 	25.7 & 	21.6 & 	21.6 & 	19.0 & 	23.9 \\
        3. Speech  &  \cellcolor{gray!15}{tcpWER} & \cellcolor{gray!15}anon+orig & \cellcolor{gray!15}17.8 & \cellcolor{gray!15}17.3\cellcolor{gray!15}& 16.7\cellcolor{gray!15}& 17.1\cellcolor{gray!15}& 19.9\cellcolor{gray!15}& \cellcolor{gray!15}17.8  & 	\cellcolor{gray!15}17.2    & 	\cellcolor{gray!15}14.2	 & \cellcolor{gray!15}13.7 & 	\cellcolor{gray!15}13.2 & 	\cellcolor{gray!15}14.8 & \cellcolor{gray!15}14.6\\
        combination & DER & anon+orig  & 41.2 & 27.8 & 16.2 & 10.6 & 10.1 & 21.2  & 	33.0 & 	22.2 & 	12.9 & 	8.4 & 	8.0	 & 16.9\\ 
        \midrule
         TSE$_{\text{attacker}}$ & \cellcolor{gray!15}{EER} & \cellcolor{gray!15}anon &  \cellcolor{gray!15}35.6 & \cellcolor{gray!15}35.8 & \cellcolor{gray!15}37.8 & \cellcolor{gray!15}35.8 & \cellcolor{gray!15}37.2 & \cellcolor{gray!15}36.4  & 	\cellcolor{gray!15}39.2 & \cellcolor{gray!15}36.2 & \cellcolor{gray!15}36.6 & \cellcolor{gray!15}35.6 & \cellcolor{gray!15}36.8 &\cellcolor{gray!15}36.9\\
     \bottomrule
     % &   1. User TSE & EER & orig & 4.2 & 4.4 & 4.2 & 4.8 & 4.8 & 4.8 \\
     %  &   & EER & anon & 31.2 & 33.0 & 31.2 & 31.4 & 31.4 & 31.2 \\
     %    2. TSA & WER & orig & 36.2 & 21.3 & 14.6 & 12.3 & 12.3 & 11.3 \\
     %     & WER & anon & 48.9 & 31.4 & 25.7 & 21.6 & 21.6 & 19.0 \\
     %    3. Speech & tcpWER & anon+orig & 30.2 & 17.2 & 14.2 & 13.7 & 13.2 & 14.8 \\
     %     combination  & DER & anon+orig & 43.2 & 33.0 & 22.2 & 12.9 & 8.4 & 8.0 \\
     %    4. TSE$_{\text{attacker}}$ & EER & anon & 39.4 & 39.2 & 36.2 & 36.6 & 35.6 & 36.8 \\
     %    \bottomrule
    \end{tabular}}
    \caption{Privacy and utility results for TSA and metrics for intermediate steps. Primary privacy and utility metrics are highlighted in gray.}
    \label{tab:summary}
\end{table*}

\subsubsection{Conformer-based TSE}
\label{ssec:conf-tse}
\vspace{-1pt}

% libri2talker\_libri2vox
%

 The conformer \cite{gulati2020conformer} architecture 
 is efficient for processing speech signals because it effectively combines two key mechanisms:
 convolutional layers and self-attention, that allow to capture  fine-grained, local patterns as well as  more global dependencies in a speech signal.
  We use a  conformer-based TSE model proposed in  \cite{liu2024target,liu2024libri2vox} that  processes the time-frequency domain short-term Fourier transform (STFT) spectrum of the  mixture audio.  The conformer-based TSE model reconstructs the real and imaginary components of the target speaker's STFT spectrum, effectively isolating their speech.
To identify and focus on the target speaker, the conformer blocks also incorporate speaker embeddings extracted from reference audio samples using a separate, pre-trained speaker encoder.
The model was trained  
 on the \textit{Libri2Mix}~\cite{cosentino2020librimix} and \textit{Libri2Vox}~\cite{liu2024libri2vox} training datasets.

\subsubsection{WeSep band-split recurrent neural network (BSRNN) TSE}
\vspace{-2pt}

The BSRNN TSE model \cite{luo2023music, yu2023tspeech} operates by explicitly dividing the audio spectrogram into multiple, distinct frequency bands. This granular approach allows for incredibly fine-grained modeling of each band's unique spectral characteristics. In this work, we use an implementation of the BSRNN model\footnote{\url{https://github.com/wenet-e2e/wesep/blob/master/wesep/models/bsrnn.py}} from the \textit{WeSep} toolkit
% \footnote{\url{https://github.com/wenet-e2e/WeSep}}
\cite{wesep_2024}.

\subsection{Evaluation models}

\subsubsection{Attack models}

The ASV system used for privacy evaluation is taken from the VoicePrivacy 2024 Challenge official evaluation setup~\cite{tomashenko2024voiceprivacy}.
It is  an ECAPA-TDNN model~\cite{thienpondt2023ecapa2} with 512 channels in the convolution
frame layers, implemented by adapting the \textit{SpeechBrain}~\cite{speechbrain} \textit{VoxCeleb} recipe to \textit{LibriSpeech train-clean-360} dataset. 
We consider a \textit{semi-informed} attacker, who has access to the anonymization system under evaluation. Using that system, the attacker anonymizes the original enrollment data so as to reduce the mismatch
with the anonymized trial data. In addition, the attacker anonymizes the training dataset
and retrains the ASV system 
% (denoted ASV anon eval ) 
on it, so that it is adapted to this specific anonymization
system.
% Anonymization is conducted on the utterance level, using the same pseudo-speaker assignment
% process as the trial data. 
For a given speaker, all enrollment utterances are used to compute an average speaker vector for enrollment.
In this privacy preservation scenario, we assume that  attackers have access to the mixture speech signal after TSA, thus they  need first to extract target speaker information from this signal. 
In particular, they  may implement TSE to this signal by using original reference embedding (as in the \textit{ignorant} attack scenario) or speaker embedding computed from the anonymized reference (as in the \textit{semi-informed} attack scenario). 

% - need to consider more advanced attack models 
% Alternaively, 
% we need to consider more advanced attack models 
\vspace{-3pt}

\subsubsection{ASR and diarization models}
\vspace{-1pt}

For utility assessment of the output TSA mixture signal, including the computation of tcpWER and DER metrics, we utilized a diarization-conditioned \textit{Whisper} for target speaker ASR
% For utility assessment of the output TSA mixture signal,
% to compute  tcpWER and DER metrics,
% we used a diarization-conditioned Whisper for target speaker automatic speech recognition
(\textit{DiCoW}) 
proposed in \cite{polok2025dicow} with
\textit{DiariZen}\footnote{\url{https://github.com/Lakoc/DiariZen}} diarization model.
\textit{DiariZen} uses local end-to-end neural diarization followed by speaker embedding clustering with \textit{pyannote}\cite{bredin2020pyannote}.
%
% This system was used to compute tcpWER and DER.
%
In addition, we employed the VoicePrivacy 2024 ASR evaluation model to evaluate the ability of the anonymization system to leave the linguistic content undistorted in the anonymized speech of the target speaker before combining it with the non-target speaker's speech.
This ASR model
% \footnote{\url{https://huggingface.co/speechbrain/asr-wav2vec2-librispeech}}
% (denoted $ASR_\text{eval}$) 
was fine-tuned on \textit{LibriSpeech-train-960} from \textit{wav2vec2-large-960h-lv60-self}
% \footnote{\url{https://huggingface.co/facebook/wav2vec2-large-960h-lv60-self}}
using \textit{SpeechBrain}.

\vspace{-4pt}

\section{Results}
\label{sec:results}

% \subsection{Research questions}
\vspace{-1pt}
\subsection{Single-speaker scenario}
\label{ref:ssec:single-speaker}
\vspace{-2pt}

% This section reports auxilry expariments for
% Privacy and utility evaluation of the
% reference speech of the target speaker  in single-speaker recordings (before mixture) to understand the performance and potential limitations of the considered anonymization algorithm of the expected results for multi speaker TSA.

This section reports auxiliary experiments for the privacy and utility evaluation of the target speaker's reference speech in single-speaker recordings (prior to mixing). These experiments aim to understand the performance and potential limitations of the considered anonymization algorithm, informing the expected results for multi-speaker TSA.
% The average results all conditions are reported
The WER and EER results for original and anonymized data are reported in Tab.~\ref{tab:single-chan}.

\vspace{-1pt}
\subsection{Performance of TSE models}
\vspace{-2pt}

% overlap	0 %	20%	40%	60%	80%	100%
% Conformer TSE	24.10	17.9	15.84	14.62	13.96	13.95
% WeSep BSRNN TSE	22.07	18.61	17.53	17.16	16.65	16.18

The SI-SDR results of the two TSE models for all overlap conditions are presented in Tab.~\ref{tab:si-sdr}. 
Both models demonstrate a decrease in SI-SDR with an increasing percentage of overlap. 
% For the non-overlap condition, the Conformer TSE model yields better results than the WeSep BSRNN, but its performance is worse for all other conditions.
The WeSep BSRNN yields better results than Conformer TSE for all conditions.

\vspace{-1pt}
\subsection{Utility and privacy TSA assessment}
\vspace{-2pt}

The summary results are reported in Tab.~\ref{tab:summary} for primary metrics (EER and tcpWER, highlighted in gray), as well as for different steps of TSA, for original and anonymized speech.
The first two lines show tcpWER and DER results for the original mixtures. 
\textit{Step} in the first column corresponds to the 
% corresponding 
step in the TSA pipeline (see Sec.~\ref{sec:tsa} and Fig.~\ref{fig:tsa}).
%
% Results for \textit{Step 1} on original data show that TSE degrades both EER and WER results w.r.t. to the original signal in the mixture (Tab.~\ref{tab:single-chan}): in average by 2.7\% abs. for EER, 11.8\% abs. for WER for Conformer TSE and  less for WeSep BSRNN TSE 1.6\% abs. for EER,
% 8.4\% abs. for WER.

Results for \textit{Step 1} on original data indicate that TSE degrades both EER and WER with respect to the original signal in the mixture (as shown in Tab.~\ref{tab:single-chan}). Specifically, Conformer TSE resulted in an average absolute degradation of 2.7\% for EER and 11.8\% for WER. WeSep BSRNN TSE showed a smaller degradation, with an average absolute change of 1.6\% for EER and 8.4\% for WER. 

% EER: 5.7-3
% WER: 17.8-6
% %
% 2.7
% 11.8
After anonymization (\textit{Step 2}), EER is similar across all conditions and closely aligns with the single-channel anonymized results (Tab.~\ref{tab:single-chan}).
 We observe a significant increase in  WER  on anonymized data.
 The primary cause of this increased WER is insertion errors, which arise from residual signals   from the non-target speaker due to imperfect separation. 
 %
 % After anonymization (Step 2), the Equal Error Rate (EER) is similar across all conditions and closely aligns with the single-channel anonymized results. We observe a significant increase in WER on anonymized data. The primary cause of this increased WER is insertion errors, which arise from residual signals from non-target speakers due to imperfect separation. 
 % Potential solutions to mitigate the increased WER include implementing improved voice activity masking for non-target speaker segments or jointly training Automatic Speech Recognition (ASR) and Target Speaker Extraction (TSE) models.
 %
%  Possible solutions to improve
% WER could be using an improved  voice activity masking of non-target speaker segments or training jointly ASR and TSE models.
To address the high WER, potential solutions include implementing improved voice activity masking for non-target speaker segments, or jointly training the ASR and TSE models.
% to better handle anonymized speech.
%
% \subsubsection{Impact of TSE}
%
tcpWER and DER calculated at \textit{Step 3} on the resulted mixture signal from TSA, show similar trends for both TSE models, while WeSep BSRNN consistently outperforms Conformer TSE. 

The privacy evaluation results, shown in the last row, were obtained under the assumption of a \textit{semi-informed} attacker who has access to anonymized enrollment and training data. Such an attacker first extracts the target speaker's voice from the mixture and then applies an ASV system to the extracted signal. In these experiments, an attacker applies the same TSE model as the user. Two main options for signal extraction were considered: using the original reference and using an anonymized reference. The evaluation focused on two key factors: how much information from the original signal of the target speaker remains in the mixture and how effective the anonymized reference is for TSE. In our experiments, the best results for both TSE were obtained when using the original speech for extraction.
Among all tested configurations, this approach yielded the best result (i.e. the strongest attack) and is reported in the table.
On average, the EER is approximately $36-37\%$, which is $12-14\%$ relatively higher than for a single-channel scenario.

\vspace{-3pt}

\section{Conclusions}
\label{sec:prior}
\vspace{-4pt}

% TSA marks a significant advancement in securing individual privacy in complex audio environments.
% This  research 
% achieved  advancement in speaker anonymization , with key findings including:
This research proposes a novel TSA framework specifically designed for privacy preservation in complex multi-speaker conversations. 
% We established an evaluation methodology to assess both the privacy and utility of anonymized speech and  analyze  inherent model limitations and challenges.
 We established an evaluation methodology to assess both the privacy and utility of anonymized speech, and to identify inherent system limitations and challenges. 
% , offering actionable solutions from both user and attacker perspectives.
Experimental results indicate a notable degradation in overall tcpWER for the best TSE model. While privacy improved and the EER was around 36-37\%, this EER, however, suggests potential for improvement in attacker capabilities or evaluation metrics. Our experiments highlighted main challenges, including the need for improved utility,
% for users,
which may require more efficient TSE jointly trained with ASR objectives.

\vfill\pagebreak

\bibliographystyle{IEEEbib}
\bibliography{strings,refs}

\begin{thebibliography}{10}

\bibitem{nautsch2019gdpr}
Andreas Nautsch, Catherine Jasserand, et~al.,
\newblock ``The {GDPR} \& speech data: Reflections of legal and technology communities, first steps towards a common understanding,''
\newblock in {\em Interspeech}, 2019, pp. 3695--3699.

\bibitem{tomashenko2020introducing}
Natalia Tomashenko, Brij Mohan~Lal Srivastava, Xin Wang, Emmanuel Vincent, Andreas Nautsch, et~al.,
\newblock ``{Introducing the {VoicePrivacy} Initiative},''
\newblock in {\em Interspeech}, 2020, pp. 1693--1697.

\bibitem{patino2020speaker}
Jose Patino, Natalia Tomashenko, et~al.,
\newblock ``Speaker anonymisation using the {McAdams} coefficient,''
\newblock in {\em Interspeech}, 2021, pp. 1099--1103.

\bibitem{mawalim22_spsc}
Candy~Olivia Mawalim, Shogo Okada, and Masashi Unoki,
\newblock ``Speaker anonymization by pitch shifting based on time-scale modification,''
\newblock in {\em 2nd Symposium on Security and Privacy in Speech Communication}, 2022, pp. 35--42.

\bibitem{gupta2020designn}
Priyanka Gupta, Gauri~P. Prajapati, Shrishti Singh, Madhu~R. Kamble, and Hemant~A. Patil,
\newblock ``Design of voice privacy system using linear prediction,''
\newblock in {\em 2020 Asia-Pacific Signal and Information Processing Association Annual Summit and Conference (APSIPA ASC)}, 2020, pp. 543--549.

\bibitem{tavi2022improving}
Lauri Tavi, Tomi Kinnunen, and Rosa~Gonz{\'a}lez Hautam{\"a}ki,
\newblock ``Improving speaker de-identification with functional data analysis of f0 trajectories,''
\newblock {\em Speech Communication}, vol. 140, pp. 1--10, 2022.

\bibitem{fang2019speaker}
Fuming Fang, Xin Wang, Junichi Yamagishi, Isao Echizen, et~al.,
\newblock ``Speaker anonymization using x-vector and neural waveform models,''
\newblock in {\em Speech Synthesis Workshop}, 2019, pp. 155--160.

\bibitem{miao2023speaker}
Xiaoxiao Miao, Xin Wang, et~al.,
\newblock ``Speaker anonymization using orthogonal {Householder} neural network,''
\newblock {\em IEEE/ACM Transactions on Audio, Speech, and Language Processing}, vol. 31, pp. 3681--3695, 2023.

\bibitem{srivastava2021}
Brij Mohan~Lal Srivastava, Mohamed Maouche, Md~Sahidullah, et~al.,
\newblock ``Privacy and utility of x-vector based speaker anonymization,''
\newblock {\em IEEE/ACM Transactions on Audio, Speech and Language Processing}, vol. 30, pp. 2383--2395, 2022.

\bibitem{champion2023}
Pierre Champion,
\newblock {\em Anonymizing speech: evaluating and designing speaker anonymization techniques},
\newblock Ph.D. thesis, Université de Lorraine, 2023.

\bibitem{yao2024musa}
Jixun Yao, Qing Wang, Pengcheng Guo, Ziqian Ning, Yuguang Yang, Yu~Pan, and Lei Xie,
\newblock ``{MUSA}: Multi-lingual speaker anonymization via serial disentanglement,''
\newblock {\em arXiv preprint arXiv:2407.11629}, 2024.

\bibitem{miao2022language}
Xiaoxiao Miao, Xin Wang, et~al.,
\newblock ``Language-independent speaker anonymization approach using self-supervised pre-trained models,''
\newblock {\em arXiv preprint arXiv:2202.13097}, 2022.

\bibitem{yao2024npu}
Jixun Yao, Nikita Kuzmin, Qing Wang, Pengcheng Guo, et~al.,
\newblock ``{NPU-NTU} system for {Voice Privacy} 2024 challenge,''
\newblock {\em arXiv preprint arXiv:2409.04173}, 2024.

\bibitem{saini2023speaker}
Shalini Saini and Nitesh Saxena,
\newblock ``Speaker anonymity and voice conversion vulnerability: A speaker recognition analysis,''
\newblock in {\em 2023 IEEE Conference on Communications and Network Security (CNS)}. IEEE, 2023, pp. 1--9.

\bibitem{webber2024voice}
Jacob~J Webber, Oliver Watts, Gustav~Eje Henter, Jennifer Williams, and Simon King,
\newblock ``Voice conversion-based privacy through adversarial information hiding,''
\newblock {\em arXiv preprint arXiv:2409.14919}, 2024.

\bibitem{miao2025benchmark}
Xiaoxiao Miao, Ruijie Tao, Chang Zeng, and Xin Wang,
\newblock ``A benchmark for multi-speaker anonymization,''
\newblock {\em IEEE Transactions on Information Forensics and Security}, 2025.

\bibitem{cosentino2020librimix}
Joris Cosentino, Manuel Pariente, Samuele Cornell, Antoine Deleforge, and Emmanuel Vincent,
\newblock ``{LibriMix}: An open-source dataset for generalizable speech separation,''
\newblock {\em arXiv preprint arXiv:2005.11262}, 2020.

\bibitem{wesep_2024}
Shuai Wang, Ke~Zhang, Shaoxiong Lin, Junjie Li, Xuefei Wang, Meng Ge, et~al.,
\newblock ``{WeSep: A Scalable and Flexible Toolkit Towards Generalizable Target Speaker Extraction},''
\newblock in {\em {Interspeech 2024}}, 2024, pp. 4273--4277.

\bibitem{watanabe20b_chime}
Shinji Watanabe, Michael Mandel, Jon Barker, Emmanuel Vincent, et~al.,
\newblock ``Chime-6 challenge: Tackling multispeaker speech recognition for unsegmented recordings,''
\newblock in {\em 6th International Workshop on Speech Processing in Everyday Environments (CHiME 2020)}, 2020, pp. 1--7.

\bibitem{polok2025dicow}
Alexander Polok, Dominik Klement, et~al.,
\newblock ``{DiCoW}: Diarization-conditioned {Whisper} for target speaker automatic speech recognition,''
\newblock {\em Computer Speech \& Language}, p. 101841, 2025.

\bibitem{MeetEval23}
Thilo von Neumann, Christoph Boeddeker, Marc Delcroix, and Reinhold Haeb-Umbach,
\newblock ``{MeetEval}: A toolkit for computation of word error rates for meeting transcription systems,''
\newblock in {\em Proc. 7th International Workshop on Speech Processing in Everyday Environments (CHiME 2023)}, 2023, pp. 27--32.

\bibitem{le2019sdr}
Jonathan Le~Roux, Scott Wisdom, Hakan Erdogan, and John~R Hershey,
\newblock ``Sdr--half-baked or well done?,''
\newblock in {\em ICASSP 2019}. IEEE, 2019, pp. 626--630.

\bibitem{tomashenko2024voiceprivacy}
Natalia Tomashenko, Xiaoxiao Miao, Pierre Champion, Sarina Meyer, et~al.,
\newblock ``The {VoicePrivacy} 2024 challenge evaluation plan,''
\newblock {\em arXiv preprint arXiv:2404.02677}, 2024.

\bibitem{liu2024libri2vox}
Yun Liu, Xuechen Liu, Xiaoxiao Miao, and Junichi Yamagishi,
\newblock ``Libri2vox dataset: Target speaker extraction with diverse speaker conditions and synthetic data,''
\newblock {\em arXiv preprint arXiv:2412.12512}, 2024.

\bibitem{gulati2020conformer}
Anmol Gulati, James Qin, Chung-Cheng Chiu, Niki Parmar, Yu~Zhang, Jiahui Yu, Wei Han, Shibo Wang, Zhengdong Zhang, Yonghui Wu, et~al.,
\newblock ``Conformer: Convolution-augmented transformer for speech recognition,''
\newblock {\em arXiv preprint arXiv:2005.08100}, 2020.

\bibitem{liu2024target}
Yun Liu, Xuechen Liu, Xiaoxiao Miao, and Junichi Yamagishi,
\newblock ``Target speaker extraction with curriculum learning,''
\newblock {\em arXiv preprint arXiv:2406.07845}, 2024.

\bibitem{luo2023music}
Yi~Luo and Jianwei Yu,
\newblock ``Music source separation with band-split rnn,''
\newblock {\em IEEE/ACM Transactions on Audio, Speech, and Language Processing}, vol. 31, pp. 1893--1901, 2023.

\bibitem{yu2023tspeech}
Jianwei Yu, Hangting Chen, et~al.,
\newblock ``{TSpeech-AI} system description to the 5th deep noise suppression ({DNS}) challenge,''
\newblock in {\em ICASSP 2023}. IEEE, 2023, pp. 1--2.

\bibitem{thienpondt2023ecapa2}
Jenthe Thienpondt and Kris Demuynck,
\newblock ``{ECAPA2}: A hybrid neural network architecture and training strategy for robust speaker embeddings,''
\newblock in {\em IEEE Automatic Speech Recognition and Understanding Workshop (ASRU)}, 2023, pp. 1--8.

\bibitem{speechbrain}
Mirco Ravanelli, Titouan Parcollet, et~al.,
\newblock ``{SpeechBrain}: A general-purpose speech toolkit,''
\newblock {\em arXiv preprint arXiv:2106.04624}, 2021.

\bibitem{bredin2020pyannote}
Herv{\'e} Bredin et~al.,
\newblock ``Pyannote. audio: neural building blocks for speaker diarization,''
\newblock in {\em ICASSP 2020}. IEEE, 2020, pp. 7124--7128.

\end{thebibliography}

\end{document}